\documentclass[aps,prb,preprint,amsmath,amssymb,superscriptaddress]{revtex4-1}

\usepackage{graphicx}
\usepackage{epstopdf}
\usepackage{amsmath}

\usepackage{color}

\begin{document}

\title{Impurity screening and stability of Fermi arcs against Coulomb and magnetic scattering in a Weyl monopnictide}

\author{Paolo Sessi} 
	\email[corresponding author: ]{sessi@physik.uni-wuerzburg.de}
	\affiliation{Physikalisches Institut, Experimentelle Physik II, 
	Universit\"{a}t W\"{u}rzburg, Am Hubland, 97074 W\"{u}rzburg, Germany}	
\author{Yan Sun}
	\affiliation{Max Planck Institute for Chemical Physics of Solids, 
	Noethnitzer Str.\ 40, 01187 Dresden, Germany}	
\author{Thomas Bathon} 
	\affiliation{Physikalisches Institut, Experimentelle Physik II, 
	Universit\"{a}t W\"{u}rzburg, Am Hubland, 97074 W\"{u}rzburg, Germany}
\author{Florian Glott} 
	\affiliation{Physikalisches Institut, Experimentelle Physik II, 
	Universit\"{a}t W\"{u}rzburg, Am Hubland, 97074 W\"{u}rzburg, Germany}
\author{Zhilin~Li} 
	\affiliation{Institute of Physics, Chinese Academy of Sciences, P.O.\ Box 603, Beijing 100190, P.\,R.\,China}
\author{Hongxiang Chen} 
	\affiliation{Institute of Physics, Chinese Academy of Sciences, P.O.\ Box 603, Beijing 100190, P.\,R.\,China}
\author{Liwei Guo} 
	\affiliation{Institute of Physics, Chinese Academy of Sciences, P.O.\ Box 603, Beijing 100190, P.\,R.\,China}
\author{Xiaolong Chen} 
	\affiliation{Institute of Physics, Chinese Academy of Sciences, P.O.\ Box 603, Beijing 100190, P.\,R.\,China}
\author{Marcus~Schmidt} 
	\affiliation{Max Planck Institute for Chemical Physics of Solids, 
	Noethnitzer Str.\ 40, 01187 Dresden, Germany}
\author{Claudia Felser} 
	\affiliation{Max Planck Institute for Chemical Physics of Solids, 
	Noethnitzer Str.\ 40, 01187 Dresden, Germany}
\author{Binghai Yan} 
	\affiliation{Max Planck Institute for Chemical Physics of Solids, 
	Noethnitzer Str.\ 40, 01187 Dresden, Germany}
\author{Matthias Bode}
	\affiliation{Physikalisches Institut, Experimentelle Physik II, 
	Universit\"{a}t W\"{u}rzburg, Am Hubland, 97074 W\"{u}rzburg, Germany}
	\affiliation{Wilhelm Conrad R{\"o}ntgen-Center for Complex Material Systems (RCCM), 
	Universit\"{a}t W\"{u}rzburg, Am Hubland, 97074 W\"{u}rzburg, Germany}

\date{\today}
\begin{abstract}
We present a quasiparticle interference study of clean and Mn surface-doped TaAs, 
a prototypical Weyl semimetal, to test the screening properties 
as well as the stability of Fermi arcs against Coulomb and magnetic scattering. 
Contrary to topological insulators, the impurities are effectively screened in Weyl semimetals. 
The adatoms significantly enhance the strength of the signal 
such that theoretical predictions on the potential impact of Fermi arcs can be unambiguously scrutinized.
Our analysis reveals the existence of three extremely short, perviously unknown scattering vectors. 
Comparison with theory traces them back to scattering events 
between large parallel segments of spin-split trivial states, strongly limiting their coherence. 
In sharp contrast to previous work [R. Batabyal {\it et al.}, Science Advances 2:e1600709] 
where similar but weaker subtle modulations were interpreted as evidence 
of quasi-particle interference originating from Femi arcs, we can safely exclude this being the case.  
Overall,  our results indicate that intra- as well as inter-Fermi arc scattering 
are strongly suppressed and may explain 
why---in spite of their complex multi-band structure---transport measurements 
show signatures of topological states in Weyl monopnictides.    
\end{abstract}
   

\maketitle

The recent discovery of materials hosting topologically protected electronic states 
opened a new era in condensed matter research \cite{HK2010,QZ2011} 
and drove the search for exotic fermionic quasiparticles 
which emerge as low-energy excitations in crystalline solids. 
Within this framework, theoretical predictions  that semimetals with broken inversion symmetry 
can host so-called Weyl fermions at points where single-degenerate 
bulk valence and conduction bands touch \cite{WFF2015,HXB2015}  
were soon verified experimentally.\cite{XBA2015,LWF2015,YLS2015,XAB2015,LXW2015}  
Weyl fermions were originally proposed by Hermann Weyl in 1929 \cite{W1929}
as chiral massless particles and have long been considered a purely abstract concept.  
Weyl fermions in condensed matter materials always appear in pairs. 
In reciprocal space, they can be seen as magnetic monopoles 
which act as sources or sinks of Berry curvature. 
At surfaces this results in the emergence of a new class 
of topologically protected electronic states that form Fermi arcs, 
which correspond to unclosed contours connecting Weyl points of opposite chirality. 
In contrast to topological insulators which require protection by time-reversal symmetry, 
Weyl semimetals do not depend on specific symmetries and are thus considered 
to be robust against any translationally invariant perturbations.  

The interest in Weyl semimetals goes well beyond their topological aspects,   
since the lack of inversion symmetry combined with strong spin-orbit coupling 
gives rise to multi-band bulk spin-split states giving rise to unconventional transport phenomena, 
such as the chiral anomaly \cite{HZL2015} and strong magnetoresistence effects,\cite{SNS2015}
making Weyl semimetals a promising platform for spintronics and magneto-electric applications. 
Fully exploiting their potential requires more than a mere identification of suitable materials,  
but calls for a detailed characterization of their physical properties. 
Within this framework, two aspects are essential: 
(i) understanding the stability of their electronic properties against perturbations and  
(ii) visualizing the transport properties of Fermi arcs and their response to scattering. 

\begin{figure*}[t]   
\centerline{\includegraphics[width=\textwidth]{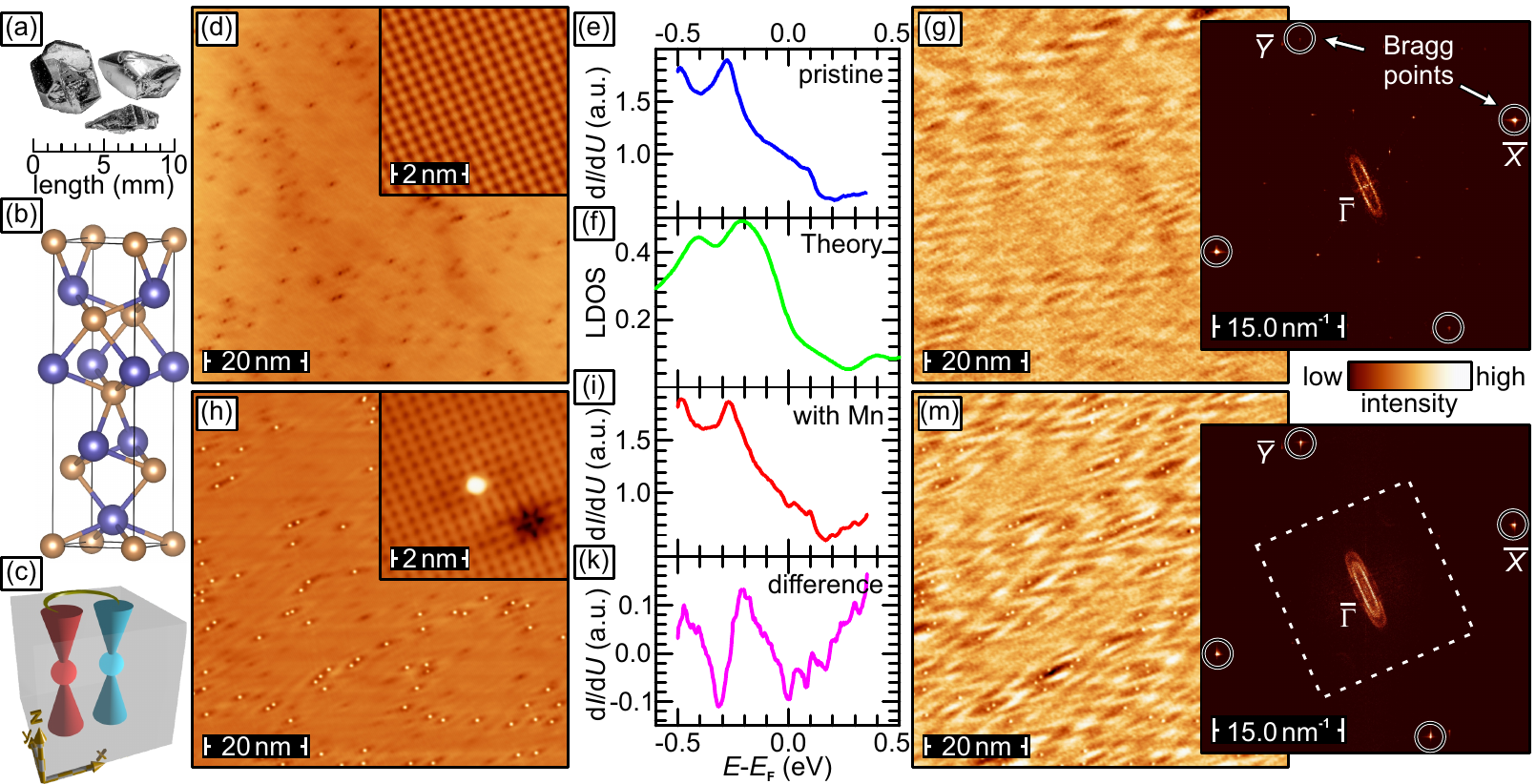}}%
\caption{{\bf Structural and electronic properties of TaAs.} 
	(a) Photography of the TaAs single crystals used in the present study. 
	(b) Crystal structure of TaAs. 
	(c) Schematic illustration of the bulk-to-surface correspondence 
	leading to the emergence of Fermi arcs. 
	(d) Topographic image of pristine TaAs (inset: atomic resolution image).
	(e) LDOS as inferred by STS on the pristine surface and 
	(f) theoretically calculated LDOS for an As-terminated surface.  
	(g) QPI map and its Fourier-transformed of clean TaAs ($E - E_{\rm F} = +50$\,meV). 
	(h) Topographic image of the Mn surface-doped TaAs.  The inset shows the Mn adsorption site. 
	(i) Experimental LDOS as measured after Mn deposition. 
	(k) Difference in the experimentally measured LDOS before and after Mn deposition. 
	(m) QPI map and its Fourier-transformed obtained on the Mn-doped surface ($E - E_{\rm F} = +50$\,meV).
	The central region within the dashed box will be analyzed in detail in Fig.\,\ref{fig:OuterDetail}.}
\label{fig:QPI}
\end{figure*}  
Here, we use TaAs as a prototypical system to investigate 
these fundamental aspects of Weyl materials at the atomic scale. 
By deliberately introducing well-defined external adatoms 
we directly probe their robustness to disorder.
We demonstrate that: (i) Weyl monopnictides can effectively screen external perturbations 
and that (ii) Fermi arcs are robust against Coulomb and magnetic scattering.
A detailed analysis of quantum coherent phenomena 
provides direct insight into the scattering mechanisms 
that affect the propagation of electronic states in Weyl materials.
Contrary to other topological materials such as topological insulators 
(TIs) \cite{RSP2009,ZCC2009,SOB2013,ZOH2014} or Dirac semimetals,\cite{JZG2014}
our results reveal the existence of a very rich scattering scenario. 
Experimental results are compared with theoretically calculated scattering patterns. 
The excellent agreement between theory and experiments 
allows for a one-to-one identification of a plethora of scattering transitions 
within the multi-band structure of Weyl semimetals.  
In particular, the improved signal-to-noise ratio reveals the existence of short scattering vectors $\mathbf q$, 
which escaped experimental detection so far.\cite{ZXB2016,IGW2016,BMA2016} 
This allows for a detailed verification of existing theoretical proposals 
dealing with intra- and inter-arc scattering events,\cite{KLW2016,MF2016,CXZ2016} 
indicating that both phenomena are strongly suppressed. 
Our results show that Fermi arcs are not susceptible to localization 
and explain why their signature can be detected in transport measurements. 

Fig.\,\ref{fig:QPI}(a) shows a photography of our bulk TaAs single crystals.\cite{LCJ2016}  
The semi-metal TaAs crystallizes in a body-centered tetragonal structure. 
As illustrated in Fig.\,\ref{fig:QPI}(b), it consists of interpenetrating Ta and As sublattices 
which are stacked on top of each other with no inversion symmetry. 
The easy cleavage plane is perpendicular to the (001) direction, 
which breaks the bonds in between two Ta and As planes.  
Theoretical calculations \cite{WFF2015,HXB2015} and photoemission data \cite{XBA2015,YLS2015} 
showed that TaAs hosts 24 Weyl points within the first surface Brillouin zone (SBZ). 
At the surface, Weyl points of opposite chirality form pairs, 
thereby giving rise to Fermi arcs as schematically sketched in Fig.\,\ref{fig:QPI}(c). 
The emergence of these arcs is a direct manifestation 
of the bulk-to-surface correspondence characterizing topological materials. 
Recently, it has been demonstrated that the extension of Fermi arcs 
is directly linked to the strength of the spin-orbit coupling.\cite{,SWY2015,LYS2016}   
Therefore, Fermi arcs in TaAs are considerably larger 
than in other Weyl monopnictides, such as NbAs, NbP and NbTa. 
This makes TaAs the optimal candidate to investigate if and how 
the open contour of Fermi arcs can give rise to quantum interference phenomena 
as proposed in Refs.\,\onlinecite{CXZ2016,MF2016,KLW2016}.

Single crystals were cleaved at $T \approx 20$\,K in ultra-high vacuum 
and immediately inserted into the scanning tunneling microscope (STM) operated at $T = 4.8$\,K. 
Scanning tunneling spectroscopy (STS) data have been obtained by lock-in technique 
by adding a bias voltage modulation at a frequency $f = 793$\,Hz. 
$\mathrm{d}I/\mathrm{d}U$ quasiparticle interference (QPI) maps were acquired 
simultaneously with topographic images in the constant-current mode
and Fourier-transformed for the assessment of scattering channels.   
In contrast to earlier studies,\cite{IGW2016,BMA2016}  
we prefer to present raw instead of symmetrized FT-QPI 
(see Ref.\,\onlinecite{Supplement} for a detailed discussion 
on the possible emergence of artifacts in symmetrized FT-QPI maps).
Figure\,\ref{fig:QPI}(d) reports an STM image of the pristine TaAs(001) surface.  
Overall, the cleaving process resulted in large atomically flat terraces 
with an extremely low defect concentration well below $10^{-4}$, confirming the high sample quality. 
The few small depressions visible on the surface were identified as As vacancies. 
The inset shows an atomically resolved image of the square lattice  
with the expected periodicity of $(3.4 \pm 0.1)$\,{\AA}.\cite{FSK1965}   
The local density of states (LDOS) has been experimentally investigated by STS [Fig.\,\ref{fig:QPI}(e)]. 
Contrary to photoemission studies, which are limited to occupied states, 
STS gives also access to unoccupied states, thus providing a complete spectroscopic characterization 
of the material above as well as below the Fermi level $E_{\rm F}$. 
The minimum visible in the $\mathrm{d}I/\mathrm{d}U$ curve 
at $E - E_{\rm F} \approx 150$\,meV highlights the semi-metallic character of TaAs 
which has been demonstrated to be of key importance 
for the occurrence of a high magnetoresistence effects.\cite{HZL2015,SNS2015} 
Since it is {\it a priori} unclear whether the cleavage plane is Ta- or As- terminated, 
we have compared the experimental tunneling spectra of Fig.\,\ref{fig:QPI}(e) 
to the theoretically calculated LDOS, obtained by projecting 
the TaAs bulk band structure onto Ta- and As-terminated (001) surfaces. 
While the Ta-terminated surface exhibits an LDOS 
which is clearly inconsistent with the experimental spectra,\cite{Supplement} 
a good agreement is evident for As-terminated TaAs(001) [compare Fig.\,\ref{fig:QPI}(e) and (f)].  

Similar to previous studies on TIs\,\cite{SRB2014,SRB2016}, additional surface defects 
were deliberately introduced to investigate the response of TaAs.  
This approach has a dual advantage: 
(i) the increased number of scattering centers significantly improves 
the signal-to-noise ratio of FT-QPI maps and 
(ii) it allows to test the response of Weyl materials to well-defined external perturbations. 
In our study we focused on Mn atoms which, because of their high spin state, 
allow to simultaneously introduce both Coulomb as well as magnetic scattering.\cite{BB2010,SBB2016} 
Their combined action significantly increases the scattering strength. 
Ultimately, this results in a substantial improvement of the signal to noise ratio 
which is of paramount importance for a clear and unambiguous identification 
of the relevant scattering mechanism in this material (see discussion in the next sections). 
Fig.\,\ref{fig:QPI}(h) reports an STM image taken subsequent to Mn deposition. 
In addition to the As vacancies already found on the pristine surface [cf.\ Fig.\,\ref{fig:QPI}(d)], 
numerous point-like protrusions with an apparent height of $\approx 70$\,pm can be recognized.   
The atomic resolution inset reveals that Mn adsorbs at the As site.\cite{Supplement} 
In Fig.\,\ref{fig:QPI}(i) we plot a tunneling spectrum measured on the Mn-doped TaAs(001) surface.  
Obviously, it is qualitatively very similar to the STS data 
obtained on the pristine surface [cf.\ Fig.\,\ref{fig:QPI}(e)].  
In particular, the peak visible in the occupied states at $E - E_{\rm F} \approx -0.3$\,eV 
indicates the absence of any shift in the band structure subsequent to deposition.
The subtle changes induced by Mn adsorption 
are highlighted by the difference spectrum shown in Fig.\,\ref{fig:QPI}(k). 
The difference in the TaAs(001) LDOS before and after Mn deposition never exceeds 6\%.
It is instructive to compare this result to similar experiments performed on TIs.
TIs exhibit a bulk band gap such that screening can only be achieved 
through two-dimensional Dirac states, resulting in strong surface band bending effects 
already at coverages below 0.01 monolayer.\cite{SRB2014} 
Although the density of states at the Fermi level of semimetals is also expected to be rather low, 
the absence of a gap and the three-dimensional character of the band structure 
results in a more effective screening of the Coulomb perturbations. 
Indeed, our results demonstrate the relative robustness of Weyl semimetals 
against Coulomb perturbations proving that they can effectively screen disorder.

In order to investigate the electronic structure of this material in more detail 
and to understand how electronic states are influenced by the presence of perturbations, 
we analyzed the standing wave pattern generated by coherent scattering at defects. 
The resulting LDOS modulations can be visualized 
by energy-resolved $\mathrm{d}I/\mathrm{d}U$ maps. 
Fig.\,\ref{fig:QPI}(g) reports a $\mathrm{d}I/\mathrm{d}U$ map 
measured at $E - E_{\rm F} = eU = +50$\,meV at exactly the same location 
as the topographic image displayed in Fig.\,\ref{fig:QPI}(d). 
These QPI maps show highly anisotropic standing wave patterns 
which are translated into reciprocal space by Fourier transformation (FT), 
thereby providing direct insight into the scattering mechanisms.  
FT-QPI maps visualize scattering vectors $\mathbf q$ 
that connect initial $\mathbf k_{\rm i}$ and final $\mathbf k_{\rm f}$ state wave vectors 
on an iso-energy contour, ${\mathbf q} = {\mathbf k_{\rm i}} - {\mathbf k_{\rm f}}$.
As shown in the inset, the Bragg points of the square As(001) surface lattice 
can clearly be recognized (highlighted by four circles). 

Figure\,\ref{fig:QPI}(m) reports a $\mathrm{d}I/\mathrm{d}U$ map 
obtained on Mn-doped TaAs(001) at the same energy as Fig.\,\ref{fig:QPI}(g).
Obviously, the additional scattering centers strongly enhances 
the intensity of the standing wave pattern as compared to the pristine case. 
As we will discuss in detail below, the strongly improved signal-to-noise ratio 
facilitates the detection of previously unresolvable scattering channels.  
Because of their implications for transport and, more generally, 
quantum coherence phenomena, it is of paramount importance to identify the states 
which, upon scattering, give rise to the observed QPI pattern.

\begin{figure*}[t]   
\centerline{\includegraphics[width=\textwidth]{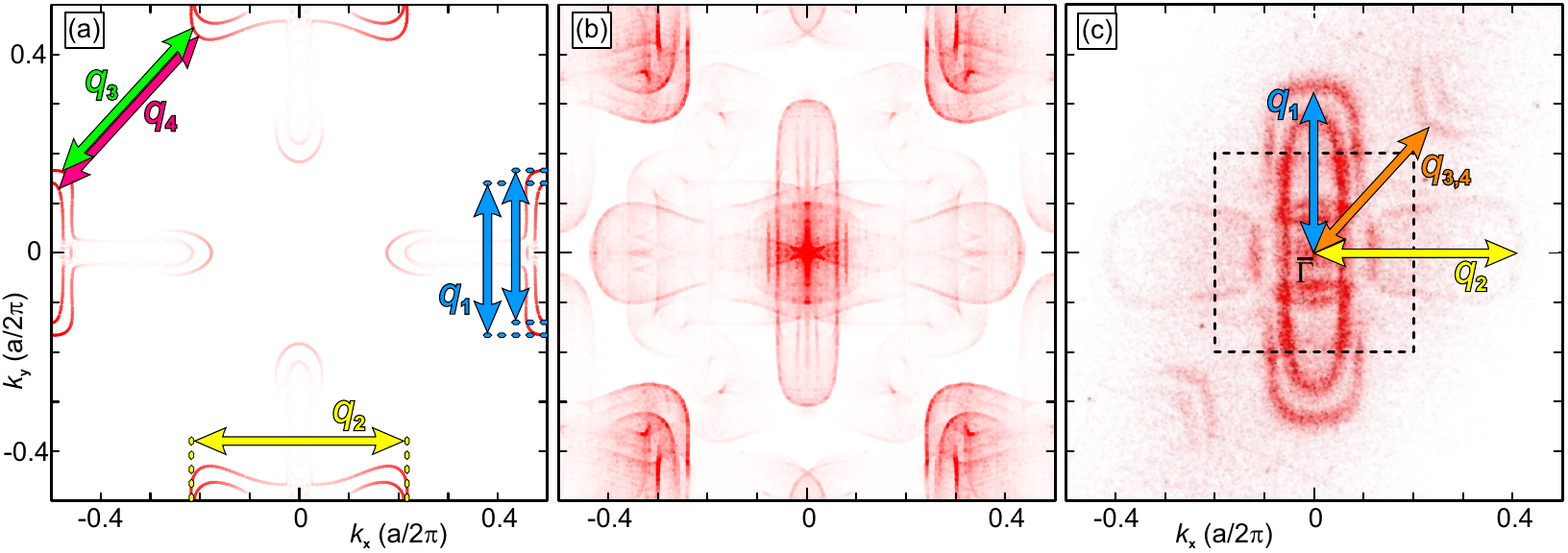}}%
	\caption{{\bf Comparison of theoretical and experimental scattering maps.}
	(a) Constant energy cut for the As-terminated surface revealing the existence of multiple energy contours. 
	(b) Theoretical ($E - E_{\rm F} = -100$\,meV) and (c) experimental FT-QPI ($E - E_{\rm F} = -90$\,meV). 
	The good agreement between theory and experiment allows to identify 
	a first subset of the dominant scattering mechanisms and to link them 
	to their corresponding initial and final states (see labelled arrows). 
	The central region within the dashed box will be analyzed in detail in Fig.\,\ref{fig:InnerDetail}.
	\label{fig:OuterDetail}}
\end{figure*}   
Fig.\,\ref{fig:OuterDetail}(a) shows a theoretical constant-energy cut 
of As-terminated TaAs(001) at $E - E_{\rm F} = -100$\,meV, 
i.e.\ at an energy where a particularly rich scattering scenario is observed. 
Bowtie-like contours are visible around the $\overline{X}$ and $\overline{Y}$ point of the SBZ. 
Closer to the center ($\overline{\Gamma}$), open spoon-like contours emerge from the background. 
These features correspond to Fermi arcs and their weak intensity signals their low surface spectral weight. 
A better comparison between experimental data and theoretically obtained scattering maps 
can be obtained by the spin-dependent joint-density-of-states (JDOS) approach [Fig.\,\ref{fig:OuterDetail}(b)]. 
Comparison with high $k$-resolution experimental results 
displayed in Fig.\,\ref{fig:OuterDetail}(c) reveals an excellent overall agreement
and will allow for the identification of a plethora of scattering vectors 
and the respective initial and final states involved in these processes. 
The only qualitative difference is the experimental observation 
of an elongated double ellipse (along the $\overline{Y}$ direction), also reported in Ref.\,\cite{IGW2016}, 
the inner part of which is absent in the theoretical FT-QPI map of Fig.\,\ref{fig:OuterDetail}(b). 
By comparing many experimental FT-QPI maps measured in an energy range 
spanning $-200$\,meV\,$\le E - E_{\rm F} \le +200$\,meV, 
we can show that this feature---which is the only one visible in the pristine case---exhibits 
essentially no dispersion.\cite{Supplement}  
This evidences that this feature is not representative of any scattering process 
but related to a set-point effect and, therefore, will not be discussed further.  

\begin{figure}[t]   
\includegraphics[width=0.7\columnwidth]{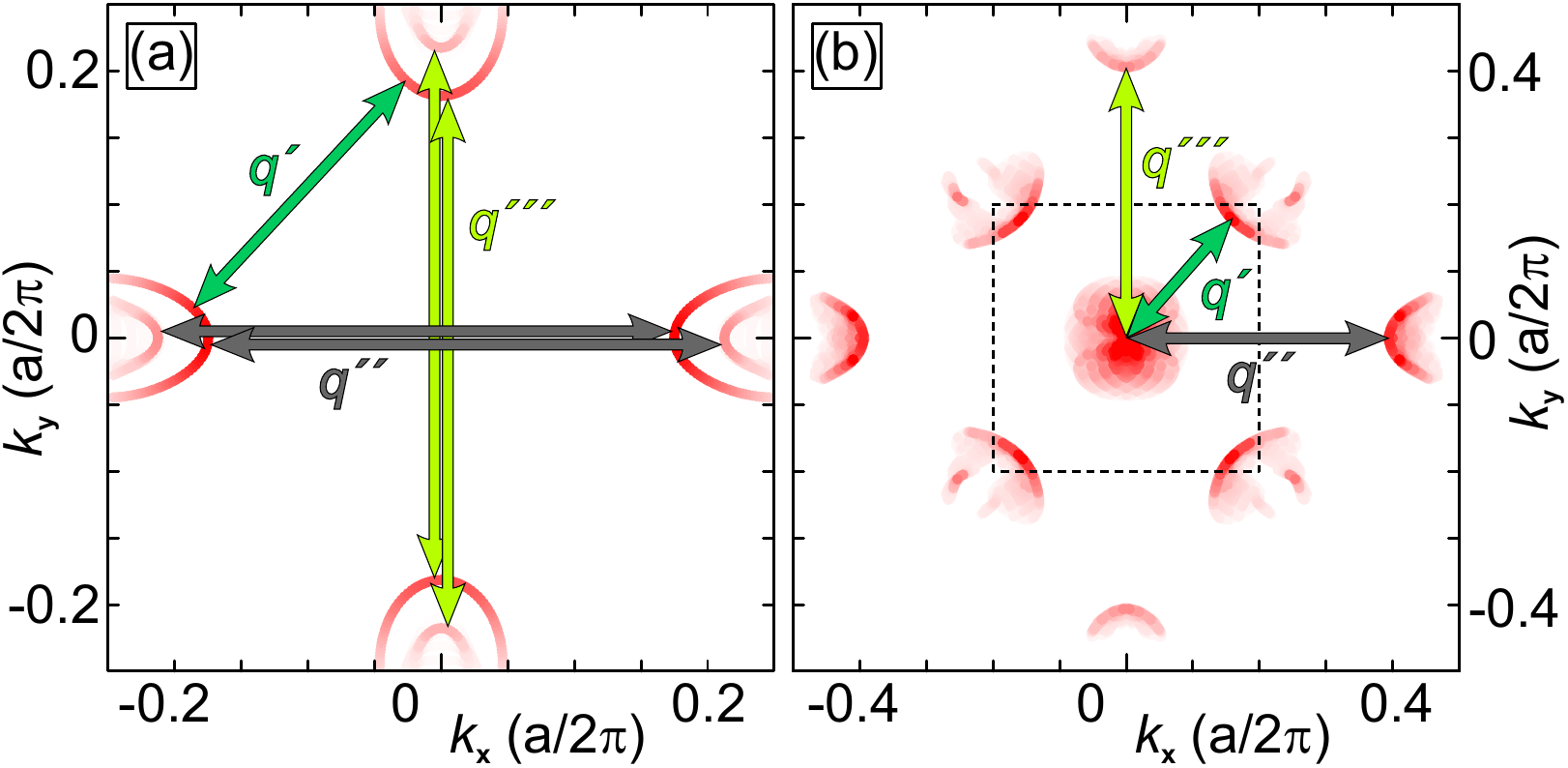}%
\caption{{\bf Fermi arc contribution to the Fermi surface and scattering vectors expected for TaAs.}
	(a) Theoretically calculated Fermi surface containing the Fermi arcs only and  
	(b) the QPI interference pattern expected from the arcs displayed in (a). 
	None of the scattering vectors is observed experimentally. 
	The region within the hatched box will be discussed in detail in Fig.\,\ref{fig:InnerDetail}.}
\label{fig:TheoryArcs}
\end{figure}  
Focusing at long scattering vectors, four main contributions 
can be identified in the experimental data of Fig.\,\ref{fig:OuterDetail}(c), 
represented by color-coded arrows labelled $\mathbf q_1$--$\mathbf q_4$. 
As schematically illustrated in Fig.\,\ref{fig:OuterDetail}(a),  
$\mathbf q_1$ corresponds to scattering vectors that connect  
the inner with outer surface contours centered at $\overline{X}$.   
Our analysis revealed that in this case intra-band backscattering is forbidden 
because of the opposite spin polarization of initial and final states. 
The same holds for $\mathbf q_2$, which involves scattering 
within energy contours centered at $\overline{Y}$.
On the other hand, as observed previously,\cite{ZXB2016,IGW2016,BMA2016} 
$\mathbf q_3$ and $\mathbf q_4$ are examples for inter-pocket scattering 
in between two contours centered around $\overline{X}$ and $\overline{Y}$, respectively.

Interestingly, additional features which up to now have escaped experimental detection 
clearly emerge from the background close to the center of Fig.\,\ref{fig:OuterDetail}(c). 
Experimentally scrutinizing this small momentum region is of fundamental importance 
for a detailed evaluation of existing theoretical proposals that suggest the emergence 
of interference phenomena originating from Fermi arcs \cite{CXZ2016,MF2016,KLW2016}. 
In order to illuminate whether the experimentally detected data contain contributions from Fermi arc, 
Fig.\,\ref{fig:TheoryArcs} reports (a)~the constant-energy contour 
and (b)~the QPI pattern that is theoretically expected 
as we restrict analysis to contributions {\em in between} surface Fermi arcs. 
It closely resembles earlier results.\cite{CXZ2016} 
The most intensive contributions, $\mathbf q'$, $\mathbf q''$, and $\mathbf q'''$, 
arise from inter-arc scattering between features along $\overline{X}$ and $\overline{Y}$ direction. 
However, these features are so weak that they only insignificantly contribute 
to the {\em total} FT-QPI pattern, which contains the sum of both, 
contributions from trivial surface-projected states 
and Fermi arcs [cf.\ Fig.\,\ref{fig:OuterDetail}(b) and (c)].
Therefore, we conclude that inter-arc scattering---although possible in principle---is so
low in intensity that it can be neglected in comparison with other scattering channels.

\begin{figure}[t]   
\includegraphics[width=0.8\columnwidth]{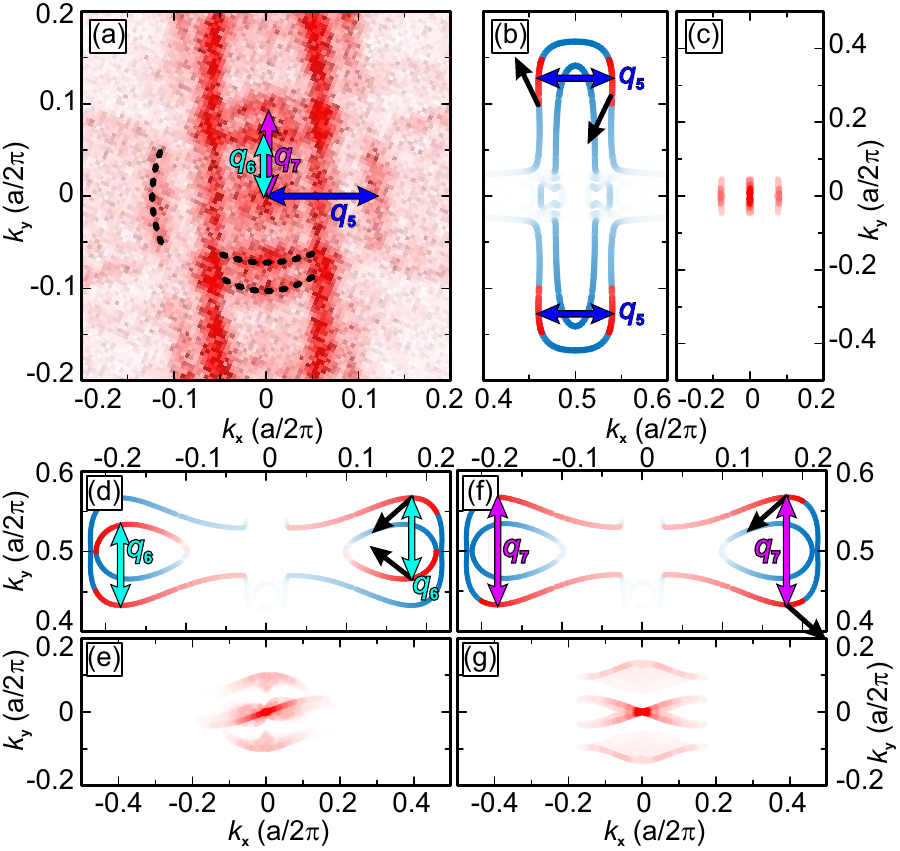}%
\caption{{\bf Short scattering vectors and theoretical analysis of their origin.} 
	(a) Detail of the FT-QPI data for short scattering vectors $\mathbf q$.
	(b)--(g) Theoretical constant-energy contours 
	(red sections contribute strongly to QPI signal) 
	and calculated QPI maps that identify the origin 
	of scattering vectors $\mathbf q_5$, $\mathbf q_6$, and $\mathbf q_7$. }
\label{fig:InnerDetail}
\end{figure}  
After excluding inter-arc scattering we want to discuss in the following 
whether intra-arc scattering significantly contributes to QPI.   
Such intra-arc scattering is the origin of the intense red-colored region in the center 
of the hatched box in Fig.\,\ref{fig:TheoryArcs}(b) 
and has been discussed in detail in Ref.\,\onlinecite{KLW2016}.  
Fig.\,\ref{fig:InnerDetail}(a) shows a high-resolution FT-QPI map 
in the central region around $\overline{\Gamma}$ 
where the background usually hampers detection of specific scattering vectors. 
Our data reveal the existence of another three, extremely short, 
well-defined scattering vectors along the $\overline{\Gamma X}$ ($\mathbf q_5$)
and along the $\overline{\Gamma Y}$ direction ($\mathbf q_6$, $\mathbf q_7$). 
Much weaker but qualitatively similar features have been recently been interpreted 
as evidence for QPI patterns originating from Fermi arcs \cite{BMA2016}.
Indeed, vectors $\mathbf q_5$--$\mathbf q_7$ are in {\em qualitative} agreement 
with recent theoretical predictions on scattering features associated 
with intra-Fermi arc scattering.\cite{KLW2016}  
{\em Quantitative} comparison reveals, however, 
that these intra-Fermi arc contributions are strongly suppressed, 
as theoretically proposed in Ref.\,\onlinecite{MF2016} (see discussion below). 

As a general remark we would like to note that an intricate data analysis, 
including symmetrization,\cite{BMA2016} may lead to conclusions 
not necessarily supported by the raw data.
For example, different proposals have been made for TaAs to predict 
which scattering vectors are visible in QPI measurements at short $q$.\cite{CXZ2016,KLW2016} 
Whereas none of these scattering vectors was directly observed in Ref.~\onlinecite{BMA2016}, 
the weak intensity visible around the $\bar \Gamma$ point of the Brillouin zone after symmetrization 
was interpreted in terms of contributions related to Fermi arcs. 
Our raw data are in obvious disagreement with this interpretation. 
Instead, as pointed out in Fig.\,\ref{fig:InnerDetail}(b)-(g), 
all three experimentally observed scattering vectors 
can consistently be explained by trivial inter-band scattering events.  
$\mathbf q_{6}$ and $\mathbf q_7$ involve the large bowtie-like contour 
centered at $\overline{Y}$ as well as the smaller contour inside.
With this respect, it is worth noticing that the existence 
of large parallel segments in the constant-energy contour 
gives rise to multiple equivalent vectors, 
which are not strictly forbidden by time-reversal symmetry 
since their spin polarization exhibits a non-vanishing projection, 
as indicated by the black arrows in Fig.\,\ref{fig:InnerDetail}(b). 
Although each one has a very low probability, their sum results in a considerable intensity 
as revealed by our theoretical data and confirmed experimentally. 
This is further corroborated by an energy-dependent analysis 
which directly linked the opening of these additional scattering channels 
to the extension parallel segments in the constant energy contour.\cite{Supplement} 

Our results show that scattering patterns on TaAs 
can consistently be explained by trivial states.  
Although our method reveals previously undetected fine details 
close to the center of the Brillouin zone, no scattering events involving Fermi arcs were found.  
These findings have important consequences for transport. 
This finding implies that the coherence of trivial states in Weyl semimetals 
is strongly limited by several scattering events. 
We expect that this results in an increased resistivity for trivial as compared to 
topologically non-trivial states which appear not to be susceptible to localization. 
In spite of the complex multi-band structure of Weyl monopnictides, 
this fundamentally different behavior may allow for the unequivocal detection 
of topological signatures in transport measurements. 
Namely, it has been predicted that in Weyl semimetals exposed to strong external magnetic fields 
the electrons are forced on closed magnetic orbits 
that consist of both bulk chiral states and surface Fermi arcs.\cite{PKV2014}
Whereas bulk scattering is explicitly considered in theoretical predictions,\cite{PKV2014} 
surface scattering has been neglected so far.     
Especially in thin films surface scattering may play a significant role 
and might be the origin of the subtle differences between experimentally observed 
and theoretically expected quantum oscillation experiments for Cd$_3$As$_2$.\cite{MNH2016}
Additionally, the existence of strong nesting suggests that highly anisotropic indirect interactions, 
such as Friedel oscillations \cite{WWL2009} or RKKY coupling,\cite{ZWL2010,SRB2016} 
may also be expected in Weyl materials.

Work at Universit\"{a}t W\"{u}rzburg has been funded 
by Deutsche Forschungsgemeinschaft within SFB 1170 ``ToCoTronics'' (project A02) 
and Priority Program ``Topological Insulators: Materials - Fundamental Properties - Devices'' 
(SPP 1666; project 1468/21-2).

\end{document}